\documentclass[aps,prd,10pt,nofootinbib,twocolumn,superscriptaddress,floatfix,notitlepage]{revtex4-1}

\usepackage{graphicx}
\usepackage{amsmath,amsfonts,amssymb}
\usepackage{color}
\usepackage[breaklinks,colorlinks,urlcolor=blue,citecolor=blue,linkcolor=magenta]{hyperref}
\usepackage{verbatim}
\usepackage{enumitem}
\usepackage{aas_macros}

\usepackage{tikz}
\usetikzlibrary{arrows.meta}

\newcommand{\Msun}{M_\odot}
\newcommand{\td}{{\rm d}}
\newcommand{\vect}[1]{\boldsymbol{#1}}

\newcommand{\be}{\begin{equation}}
\newcommand{\ee}{\end{equation}}
\newcommand{\bea}{\begin{equation} \begin{aligned}}
\newcommand{\eea}{\end{aligned} \end{equation}}

\begin{document}

\title{FLUMEN: Neural Emulator of an Advanced Stochastic Weak Lensing Model}

\author{Galymzhan Baltabay}
\email{galymzhan.baltabay@kbfi.ee}
\affiliation{Laboratory of High Energy and Computational Physics, KBFI, R{\"a}vala 10, Tallinn, 10143, Estonia}
\affiliation{Department of Cybernetics, Tallinn University of Technology, Akadeemia tee 21, 12618 Tallinn, Estonia}

\author{Ville Vaskonen}
\email{ville.vaskonen@kbfi.ee}
\affiliation{Laboratory of High Energy and Computational Physics, KBFI, R{\"a}vala 10, Tallinn, 10143, Estonia}

\begin{abstract}
Weak gravitational lensing by intervening structure magnifies distant sources and induces scatter in their luminosity distances. We develop an improved stochastic model for the distribution of weak lensing magnifications, $\td P/\td\mu$, that resolves individual lenses down to $10^7\,M_\odot$ and, besides field halos and filaments, includes subhalos and large-scale clustering of the lens population. Both extensions broaden the distribution, and the resulting scatter in the luminosity distances agrees with the halo-model prediction from the nonlinear matter power spectrum and exceeds that obtained from $N$-body ray tracing, which we attribute to the finite mass resolution and box size of the simulations. Building on this model, we develop \texttt{FLUMEN}, a conditional normalizing-flow emulator that predicts $\td P/\td\mu$ as a function of the source redshift up to $z_s=12$ and six cosmological parameters $\{h,\Omega_M,\Omega_B,\sigma_8,n_s,z_{\rm eq}\}$. The emulator enforces the empty-beam cutoff and the fold-caustic power-law tail analytically, and reproduces the Monte Carlo distributions on held-out cosmologies with a median Kullback--Leibler divergence of $0.0048$ at negligible cost. Our results provide a fast and flexible framework for incorporating the full magnification distribution into cosmological analyses of standard sirens, supernovae, and other point sources.
\end{abstract}

\maketitle

\section{Introduction}

The observed luminosities and distances of distant sources are affected by intervening structures through gravitational lensing. While the mean lensing effect is determined by the average matter distribution, the stochastic magnifications contain additional information about the distribution and evolution of cosmic structure. This makes the weak lensing magnification distribution a useful probe in a broad range of cosmological analyses, including studies based on supernovae~\cite{Hamana:1999rk,Wang:2004ax,Dodelson:2005zt,Quartin:2013moa,Castro:2014oja,Zhai:2019sgi,DES:2024lto}, quasars~\cite{Hamana:1999bf}, and gravitational wave (GW) sources~\cite{Congedo:2018wfn,Mpetha:2024xiu,Vaskonen:2026ubd}. One particularly promising application is provided by GW standard sirens, namely GW signals from compact binary mergers with an identified electromagnetic counterpart. For such signals, the luminosity distance is inferred directly from the GW waveform amplitude without a distance-ladder calibration. Ref.~\cite{Vaskonen:2026ubd} showed that the lensing-induced scatter in the standard-siren distance-redshift relation can be used to constrain $\sigma_8$, complementing standard-siren measurements of the Hubble constant $H_0$ and the matter abundance $\Omega_M$. A population of $\mathcal{O}(100)$ bright sirens from third-generation detectors could provide a competitive measurement of $\sigma_8$ through this effect.

Exploiting the weak lensing scatter for cosmological inference requires an accurate model of the magnification probability distribution $\td P/\td \mu$. One route is ray tracing through the density field of cosmological $N$-body simulations~\cite{Holz:1997ic,Holz:2004xx,Takahashi:2011qd,Takahashi:2017hjr}. To accelerate inference with such simulations, machine-learning emulators such as \texttt{ACE-Lensing}~\cite{Turker:2025rdt} have been trained directly on $N$-body ray-tracing catalogs. However, generating the training data still requires a large suite of computationally expensive $N$-body simulations. The high computational cost in turn limits the mass resolution that can be achieved, imposing a high halo mass cutoff and smoothing out small-scale substructure. In particular, the \texttt{ACE-Lensing} emulator was trained on $N$-body simulations with a particle mass resolution of $\mathcal{O}(10^{9}\,\Msun)$.

The stochastic approach of Refs.~\cite{Kainulainen:2009dw,Kainulainen:2010at,Kainulainen:2011zx}, implemented in \texttt{C++} in Ref.~\cite{Vaskonen:2026ubd}, instead builds $\td P/\td\mu$ by directly summing the contributions of a Monte Carlo-realized population of discrete lenses down to $10^7\,\Msun$, resolving magnifications that $N$-body ray tracing misses at a fraction of the computational cost. Even so, evaluating $dP/d\mu$ at the many points of a cosmological parameter space required for Markov chain Monte Carlo (MCMC) inference remains expensive, since each evaluation requires a fresh realization of the lens population. This motivates building a fast, differentiable neural emulator directly for the stochastic lens model.

In this work we extend the stochastic lensing model of Ref.~\cite{Vaskonen:2026ubd} in two ways. First, we include the previously unmodeled contribution of subhalos, sampling the substructure of every lensing halo down to a fixed mass floor and removing the corresponding mass from the smooth halo profile so that the total halo mass is conserved. Second, we replace the independent Poisson realization of the halo and filament counts with a physically motivated treatment of their large-scale clustering, modulating the lens population along the line of sight, together with the sub-threshold background, by a stochastic realization of the long-wavelength density field. Both extensions broaden the magnification distribution, and we show that the resulting scatter in the luminosity distance agrees with the halo-model prediction from the nonlinear matter power spectrum, while $N$-body-based estimates fall below it because of their finite mass resolution and box size. 

Building on the improved stochastic model, we train \texttt{FLUMEN} (Fast Lensing Universal Magnification Emulator with Normalizing flows), a conditional normalizing-flow emulator of $\td P/\td\mu$ as a function of the source redshift $z_s\le12$ and the cosmological parameters $\vect\Theta=\{h,\Omega_M,\Omega_B,\sigma_8,n_s,z_{\rm eq}\}$, which builds in the empty-beam cutoff and the fold-caustic tail analytically. It reaches an accuracy comparable to that of existing emulators of the magnification distribution while incorporating the extended physical model and resolving lensing structures to substantially lower masses.\footnote{The emulator is publicly available at \url{https://github.com/baltabaygal/flumen}.}

\section{Weak lensing}
\label{sec:lensing}

Weak gravitational lensing by structures near the line of sight changes the solid angle subtended by a source, so that the observed flux of an electromagnetic source, or the squared strain amplitude of a GW source, is magnified by a factor $\mu$ relative to its value in a homogeneous and isotropic universe, while the observed redshift is unaffected. Correspondingly, the apparent luminosity distance inferred at fixed redshift is
\be
    D_L(z) = \frac{\tilde{D}_L(z)}{\sqrt{\mu}}\,,
\ee
where the tilde denotes the luminosity distance-redshift relation in the
homogeneous and isotropic universe.

The magnification of an individual source cannot be measured directly. It can be partly corrected for using the observed foreground structure along the line of sight, but such delensing requires deep foreground data and leaves a residual scatter from unresolved low-mass structure~\cite{Dalal:2002wh,Jonsson:2006vc,Shapiro:2009sr,Hilbert:2010am,SDSS:2013wlv,Wu:2022vrq}. Inferring $\tilde D_L(z)$, and hence the cosmological parameters, from a set of distance measurements therefore requires treating $\mu$ as a random variable described by the probability distribution function (PDF) $\td P/\td\mu$. This distribution depends on the abundance and clustering of the intervening structures, so that the lensing scatter is not only a nuisance for the distance-redshift relation but also a cosmological probe in its own right, provided that $\td P/\td\mu$ can be accurately modeled.

In the following we briefly describe the stochastic approach that we use for the computation of $\td P/\td\mu$, focusing on how we extend the lens modeling beyond that of Ref.~\cite{Vaskonen:2026ubd}.

\subsection{Stochastic approach}

In terms of the convergence $\kappa$ and the shear $\gamma = \sqrt{\gamma_1^2+\gamma_2^2}$, the lensing magnification is given by
\be \label{eq:mueq}
    \mu = \frac{1}{(1-\kappa)^2 - \gamma^2} \,.
\ee 
To leading order in the lens potentials, the convergence and shear of the different structures along the line of sight are additive,\footnote{The additivity is accurate both when all lenses along the line of sight are weak and when the lensing is dominated by a single lens. The latter is the regime that produces the high-magnification tail.} so that
\bea
    &\kappa = \sum_j \kappa_j(\vect\theta_j)\,, \\
    &\gamma_1 = \sum_j \gamma_j(\vect\theta_j)\cos 2\phi_j\,, \\
    &\gamma_2 = \sum_j \gamma_j(\vect\theta_j)\sin 2\phi_j\,,
\eea
where $\phi_j$ is the polar angle of the position of lens $j$ in the lens plane, measured in a coordinate system common to all lenses, and $\vect\theta_j$ collects the lens redshift $z_j$, its projected distance $r_j$ from the line of sight to the source, and the parameters describing the lens profile and orientation.

We include lensing contributions from field halos, filaments, and subhalos, together with a stochastic background accounting for the halos too numerous to be sampled individually. Field halos and filaments are modeled as in Ref.~\cite{Vaskonen:2026ubd}, with two extensions introduced in this work: the large-scale clustering of the lens population, correlated with the long-wavelength density field, and the contribution of subhalos.

\subsubsection{Clustering}

Long-wavelength linear density perturbations modulate the spatial distribution of halos. The resulting change in the halo number density is governed by the halo bias $b(M,z)$, which increases with halo mass $M$ because massive  halos originate from increasingly rare peaks of the primordial density field. 

We estimate the halo clustering using the one-dimensional density field $\delta_{\rm 1D}(z) \equiv \delta_{\rm 1D}(d_c(z))$, obtained by smoothing the three-dimensional density field isotropically with a window of radius $R_{\rm s}$ before projecting onto the line of sight. Its power spectrum is\footnote{This generalizes the pencil-beam construction of Ref.~\cite{1991ApJ...379..482K}, whose $P_{\rm 1D}(k_\parallel)=\int_{k_\parallel}^\infty \td k\,k\,P(k)/(2\pi)$ is recovered in the $R_{\rm s}\to0$ limit.}
\be
    P_{\rm 1D}(k_\parallel) = \frac{1}{2\pi} \int_{k_\parallel}^\infty \!\td k \, k \, P(k) \tilde{W}^2(R_{\rm s} k) \,.
\ee
The smoothing suppresses the contribution of scales smaller than $R_{\rm s}$ both along and perpendicular to the line of sight. We use the real-space top-hat window function and fix $R_s = 20\,$Mpc.

On a periodic path of length $L$, padded beyond the source distance $d_c(z_s)$ so that the periodicity does not spuriously correlate structure near the observer with that near the source, the 1D density field is given by
\be
    \delta_{\rm 1D}(z) = \frac{1}{L} \sum_{n\geq 1} \tilde\delta_n e^{i k_n d_c(z)} + {\rm c.c.} \,, \quad k_n= \frac{2\pi n}{L} \,,
\ee
where $d_c(z)$ denotes the comoving distance and mode orthogonality gives $\langle\tilde\delta_n\tilde\delta_{n'}^*\rangle=LP_{\rm 1D}(k_n)\delta_{nn'}$. The real and imaginary parts of  $\tilde\delta_n$ are independent Gaussian random variables with equal variance 
\be
    \sigma_n^2 = \frac{L}{2} P_{\rm 1D}(k_n) \,. 
\ee    
Given a realization of the 1D density field $\delta_{\rm 1D}(z)$, the number density of field halos along the line of sight is
\be \label{eq:n}
    \frac{\td n(z)}{\td \ln M} = \lambda(M,z) \frac{\td \bar{n}(z)}{\td \ln M} \,,
\ee
where \footnote{To ensure positivity of $n$ we use the lognormal model of Ref.~\cite{1991MNRAS.248....1C} rather than the linear one $\lambda(M,z) = 1+\tilde{b}(M,z)\,\delta_{\rm 1D}(z)$.}
\be \label{eq:lambda}
    \lambda(M,z) = \exp\!\left[ \tilde{b}(M,z) \delta_{\rm 1D}(z) - \tfrac12 \tilde{b}(M,z)^2 \sigma_{\rm 1D}^2 \right] \,.
\ee
The second term in the exponent in $\lambda(M,z)$ with 
\be
    \sigma_{\rm 1D} ^ 2 = \frac{1}{\pi} \int_0 ^ \infty \td k_\parallel P_{\rm 1D}(k_\parallel) = \frac{4}{L^2} \sum_{n\geq 1} \sigma_n ^ 2
\ee
ensures that the ensemble average is normalized to unity, $\langle \lambda(M,z)\rangle_{\rm ens} = 1$. The power spectrum $P_{\rm 1D}(k_\parallel)$ is computed from the present-day linear matter power spectrum $P(k)$, and the redshift dependence of the clustering is carried entirely by $\tilde b(M,z) = D(z) b(M,z)$, which combines the halo bias $b(M,z)$ with the linear growth factor $D(z)$ of density perturbations.
For the bias of field halos and filaments, we use the peak-background-split expression derived from the collapse barrier~\cite{Sheth:1999mn},
\be \label{eq:bias}
    b(M,z) = 1 + \frac{q\nu^2 - 1}{\delta_c} + \frac{2p}{\delta_c\left[1+(q\nu^2)^p\right]} \,,
\ee
where $\nu = \delta_c(z)/\sigma(M)$. In each case, we evaluate Eq.\eqref{eq:bias} using the same barrier parameters as those adopted for the corresponding mean abundance, discussed in the next subsection. Thus, the bias is obtained from the peak-background split of the barrier defining each population: $(p,q)=(0.3,\,0.8)$ for field halos, and $(p,q)=(0,\,0.7)$ for filaments. This gives a $10\text{--}20\%$ lower bias for filaments relative to halos of the same mass, reflecting their weaker large-scale clustering~\cite{Yan:2011ux}.

\begin{figure}
    \centering
    \includegraphics[width=1.0\linewidth]{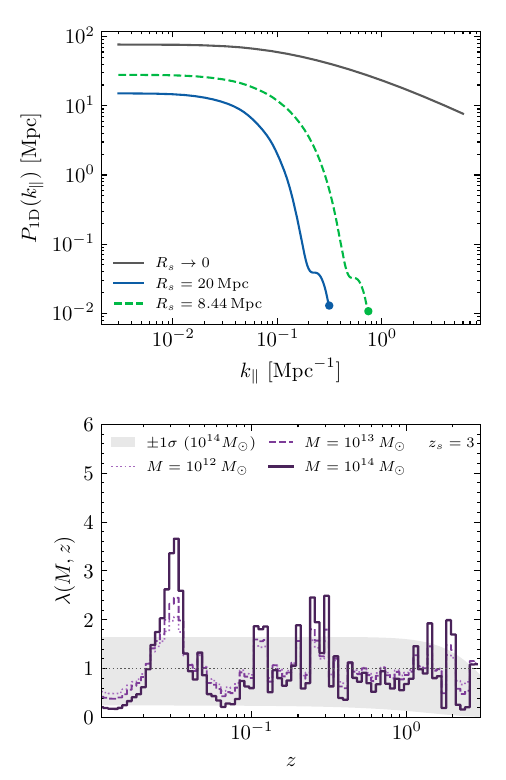}
    \caption{
    \textbf{Top:} Power spectrum of the one-dimensional density field
    $\delta_{\rm 1D}$ obtained with the top-hat window function. The dots mark the mode cutoff $k_{\rm max} = 2\pi/R_s$ of the discrete realization. \textbf{Bottom:} The count modulation $\lambda(M,z)$ for different halo masses for a source at $z_s=3$ for one realization of the underlying smoothed 1D density field. The gray band indicates the $1\sigma$ range of $\lambda$.}
    \label{fig:clustering_field}
\end{figure}

The power spectrum of $\delta_{\rm 1D}$ and example realizations of the count modulation $\lambda$ are shown in Fig.~\ref{fig:clustering_field}.

\subsubsection{Field halos and filaments}

The mean abundances of field-halos and filaments are given by the excursion set model~\cite{Sheth:1999mn,Shen:2005wd},
\be \label{eq:hmf}
    \frac{\td\bar n}{\td\ln M} = \frac{\rho_m}{M}\,A\left[1+(q\nu^2)^{-p}\right]
    \sqrt{\frac{q\nu^2}{2\pi}}\,e^{-q\nu^2/2}\left|\frac{\td\ln\nu^2}{\td\ln M}\right|,
\ee
where $\nu = \delta_c(z)/\sigma(M)$, $\rho_m$ is the mean comoving matter density, and $A = [1+2^{-p}\Gamma(\tfrac12-p)/\sqrt\pi]^{-1}$. The parameters $p$ and $q$ characterize the first-crossing distribution obtained from random walks with a moving barrier. Consistent with the bias~\eqref{eq:bias}, we use $(p,q) = (0.3,\,0.8)$ for field halos and $(p,q) = (0,\,0.7)$ for filaments.

Each field halo is assigned an NFW profile with the concentration-mass relation of Ref.~\cite{Ludlow:2016ifl},\footnote{We use the fit of Ref.~\cite{Ludlow:2016ifl}, which is calibrated over a wider range of masses and redshifts than the fit of Ref.~\cite{Dutton:2014xda} used in Ref.~\cite{Vaskonen:2026ubd}, holding it fixed at its $z=7$ value at higher redshifts. We have checked that the two give nearly identical results.} projected with an elliptical distortion whose axis ratio follows the $N$-body calibration of Ref.~\cite{Allgood:2005eu}, $s = 0.54\,(M/M_*(z))^{-0.05}$, $\epsilon = (1-s)/(1+s)$, where $M_*(z)$ is the characteristic collapse mass. The convergence and shear of each halo follow from the closed-form NFW lensing functions given in Ref.~\cite{Wright:1999jc}. Filaments are modeled as uniform-density cylinders with random orientations, as in Ref.~\cite{Vaskonen:2026ubd}, with radius $r_{\rm F} = 1\,{\rm Mpc} (M/10^{14}\Msun)^{1/3}$, length $L_{\rm F} = 20\,{\rm Mpc} (M/10^{14}\Msun)^{1/3}$, and mean density $14.4\rho_{\rm c}$.

The expected differential number of lenses at proper projected distance $r$ from the line of sight is\footnote{We use units with $c = G = 1$.}
\be \label{eq:dN}
    \frac{\td \bar N}{\td z \, \td \ln M \, \td r} =
    \frac{2\pi \, (1+z)^2 r}{H(z)} \, \frac{\td \bar{n}}{\td \ln M} \,.
\ee
We generate individually only the halos and filaments whose convergence exceeds a threshold value $\kappa_{\rm thr}$. We choose this threshold such that integrating Eq.~\eqref{eq:dN} over the full mass range, redshifts up to the source redshift $z_s$, and the radii $\{r:\kappa>\kappa_{\rm thr}\}$ yields an expected number of individually generated field halos of $\bar N_l = 100$. The remaining sub-threshold population occupies the complementary region, $A_W(M,z) = \{ r :  \kappa_{\rm NFW}(r,M,z) < \kappa_{\rm thr} \}$, and we discuss their contribution below. The number of field halos and filaments in each mass and redshift cell is Poisson distributed with the mean of Eq.~\eqref{eq:dN} modulated by the clustering factor $\lambda$ of Eq.~\eqref{eq:lambda}.

\subsubsection{Sub-threshold contribution}

The number of halos with $\kappa < \kappa_{\rm thr}$ is too large to be explicitly sampled, but their collective effect can be non-negligible. As in Ref.~\cite{Vaskonen:2026ubd}, we include them through a stochastic background convergence $\kappa_W$ added to every realization. In~\cite{Vaskonen:2026ubd}, $\kappa_W$ was drawn from an unconditional Gaussian distribution. Here we instead construct the Gaussian conditionally on the same realization of $\delta_{\rm 1D}$ that modulates the discrete lens counts, so that the resolved lenses and the sub-threshold background fluctuate coherently with the large-scale density perturbations. We compute the variance $\sigma_W^2$ of $\kappa_W$ below. We include only the convergence contribution from field halos and neglect the shear contribution from all sub-threshold structures.

Conditional on a realization of $\delta_{\rm 1D}$, the halo counts form a Poisson process with the modulated density of Eq.~\eqref{eq:n}, and Campbell's theorem gives the conditional mean and variance of the summed sub-threshold convergence as
\bea \label{eq:weakcond}
    &{\rm E}\!\left[ \kappa_W | \delta_{\rm 1D} \right] =
    \int_0^{z_s} \!\td z \int \td \ln M \, \left[\lambda(M,z) - 1\right] w_1(M,z) \,,  \\
    &{\rm Var}\!\left[ \kappa_W | \delta_{\rm 1D} \right] =
    \int_0^{z_s} \!\td z \int \td \ln M \, \lambda(M,z)\, w_2(M,z) \,,
\eea
where the moments of the sub-threshold convergence per redshift and mass are
\be \label{eq:wk}
    w_n(M,z) = \int_{A_W} \td r \,
    \frac{\td \bar N}{\td z \, \td \ln M \, \td r}  \kappa_{\rm NFW}(r)^n \,.
\ee
The mean of $\kappa_W$ vanishes, $\langle {\rm E}\!\left[ \kappa_W | \delta_{\rm 1D} \right] \rangle_{\rm ens} = 0$, and the variance of $\kappa_W$ can be split into one-halo (shot-noise) and two-halo (correlation) contributions,
\be \label{eq:sigmaW}
    \sigma_W^2 = \sigma_{\rm 1h}^2 + \sigma_{\rm 2h}^2 \,,
\ee
with
\be \label{eq:sigmashot}
    \sigma_{\rm 1h}^2 = \big\langle {\rm Var}[\kappa_W|\delta_{\rm 1D}] \big\rangle_{\rm ens} = \int_0^{z_s} \td z \int \td \ln M \; w_2(M,z)
\ee
and\footnote{In the linear limit $|\tilde b \tilde b' \xi_{\rm 1D}| \ll 1$ the two-halo contribution reduces to the familiar form $\sigma_{\rm 2h}^2 \simeq \int \td z \int \td z' \, B(z) B(z') \, \xi_{\rm 1D}$ with the bias-weighted kernel $B(z) = \int \td \ln M \, \tilde b(M,z) \, w_1(M,z)$.\cite{Seljak:2000gq}}
\bea \label{eq:sigma2h}
    &\sigma_{\rm 2h}^2 = \big\langle {\rm E}[\kappa_W|\delta_{\rm 1D}]^2 \big\rangle_{\rm ens} \\
    &= \int_0^{z_s} \!\!\td z \td z' \int \!\td \ln M \td \ln M' w_1 w_1' \left[ e^{\tilde b \tilde b' \xi_{\rm 1D}} - 1 \right] ,
\eea
where we used $\langle \lambda \lambda' \rangle_{\rm ens} = \exp[ \tilde b \tilde b' \xi_{\rm 1D}\big(|d_c - d_c'|\big) ]$ and defined $\xi_{\rm 1D}$ as the correlation function of $\delta_{\rm 1D}$:
\bea \label{eq:xi1D}
    \xi_{\rm 1D}(\Delta d_c) &= \frac{1}{\pi} \int_0^\infty \td k_\parallel \, P_{\rm 1D}(k_\parallel) \cos(k_\parallel \Delta d_c) \\
    &= \frac{4}{L^2} \sum_{n\geq 1} \sigma_n^2 \cos(k_n \Delta d_c) \,.
\eea

\begin{figure}
    \centering
    \includegraphics[width=1.0\columnwidth]{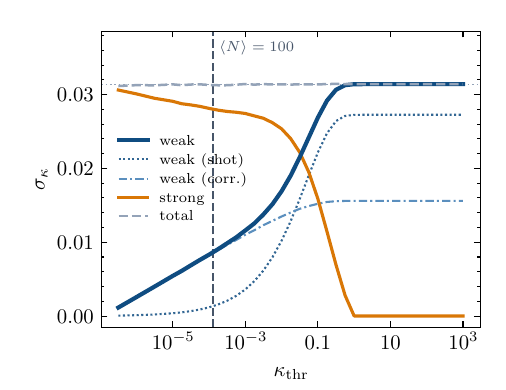}
    \caption{Partition of the convergence fluctuations at $z_s=1$ between the sub-threshold background (weak), the explicitly sampled lenses (strong), and their sum, as a function of the threshold $\kappa_{\rm thr}$.}
    \label{fig:kappa_convergence}
\end{figure}

Figure~\ref{fig:kappa_convergence} shows the resulting partition of the convergence standard deviation at $z_s=1$ between the sub-threshold background $\sigma_W$ and the explicitly sampled lenses as a function of $\kappa_{\rm thr}$. The total is independent of $\kappa_{\rm thr}$, confirming that the two treatments are consistent at the level of the variance. The threshold only controls how much of the non-Gaussian information is kept in the discrete lenses. At the adopted threshold, the background carries about a quarter of $\sigma_\kappa$ and is dominated by the correlated two-halo term rather than by shot noise, which is the contribution that the unconditional background of Ref.~\cite{Vaskonen:2026ubd} did not capture.

\subsubsection{Subhalos}

\begin{figure}
    \centering
    \includegraphics[width=1.0\linewidth]{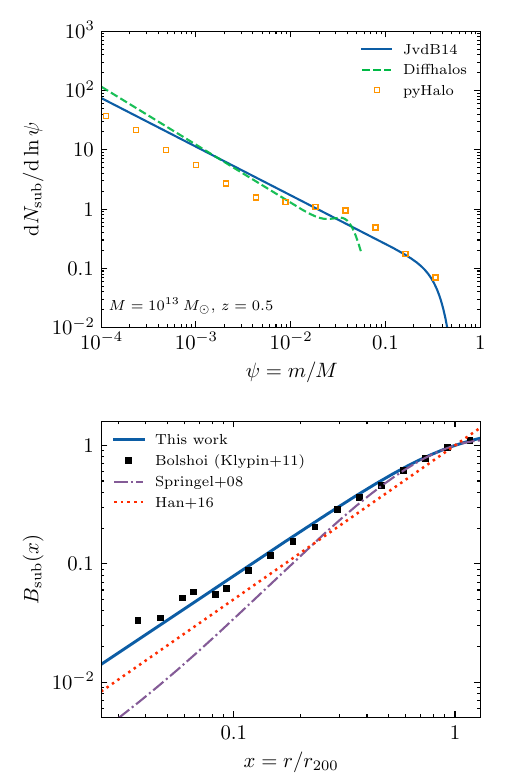}
    \caption{
    \textbf{Top:} The evolved subhalo mass function, Eq.~\eqref{eq:shmf}, for a host of mass $M = 10^{13}\Msun$ at $z=0.5$, normalized by the bound fraction. The solid curve is adapted from Ref.~\cite{Jiang:2014nsa}. For comparison we overlay the results of \texttt{Diffhalos}~\cite{Zacharegkas:2026zzy} and  \texttt{pyHalo}~\cite{Gilman:2021sdr}. \textbf{Bottom:} The radial bias function of the subhalo distribution. The adopted function shown by the solid curve is fit to the Bolshoi simulations~\cite{Klypin:2010qw} (squares). For comparison we also plot the analytic curves from Ref.~\cite{Springel:2008cc} and Ref.~\cite{Han:2015pua}.}
    \label{fig:shmf}
\end{figure}

Each lensing halo of mass $M$ is split into a reduced smooth component and a population of discrete subhalos,
\be \label{eq:reducedhost}
    \kappa_{\rm halo} = \kappa_{\rm NFW}\big(M - \textstyle\sum_i m_i\big) + \sum_i \kappa_{\rm NFW}(m_i) \,.
\ee
The subhalo masses $m$ are drawn from the evolved subhalo mass function of Ref.~\cite{Jiang:2014nsa}, a Schechter-like fit to the output of semi-analytic merger-tree models calibrated on $N$-body simulations:
\be \label{eq:shmf}
    \frac{\td N_{\rm sub}}{\td \ln\psi} = \tilde\gamma\, \psi^{\alpha} e^{-\beta \psi^{\omega}} \,, \qquad \psi \equiv \frac{m}{M} \leq 1 \,,
\ee
with $(\alpha,\beta,\omega) = (-0.78,\,50,\,4)$.\footnote{We include only first-order subhalos, i.e.\ those bound directly to the host rather than to another subhalo. Sub-subhalos are absorbed in the smooth profile of their parent.} The normalization $\tilde\gamma$ is fixed by the bound substructure mass fraction $f_{\rm s} = f_{\rm s}(M,z)$ as
\be \label{eq:fsnorm}
    \tilde\gamma = \frac{f_{\rm s} \, \omega \, \beta^s}{\Gamma\!\left(s,\beta\psi_{\rm res}^{\omega}\right) - \Gamma\!\left(s,\beta\right)} , \quad s = \frac{1+\alpha}{\omega} \,,
\ee
where $\psi_{\rm res}=10^{-4}$ and $\Gamma(s,x)$ is the upper incomplete gamma function. Fitting the semi-analytic merger-tree models calibrated on $N$-body simulations, Ref.~\cite{Jiang:2014nsa} found that the mass fraction is well described by the dynamical age of the host, 
\be
    f_{\rm s}(M,z) = 0.3563\, N_\tau(M,z)^{-0.6} - 0.075 \,, 
\ee
where
\be
    N_\tau(M,z) = \!\int_{z}^{z_f} \!\frac{\td z' \, \tau_{\rm dyn}(z')^{-1}}{(1+z') H(z')} \,, \quad  \tau_{\rm dyn}(z) = \sqrt{\frac{3\pi}{16 \bar{\rho}_h(z)}} \,,
\ee
counts the number of dynamical timescales $\tau_{\rm dyn}$ elapsed since the halo formation redshift $z_f$. Here $\bar{\rho}_h(z) = (18\pi^2 + 82\,d - 39\,d^2) \rho_c(z)$, with $d \equiv \Omega_M(z)-1$ and the critical density $\rho_c(z)$, is the mean density of a virialized halo at redshift $z$~\cite{Bryan:1997dn}. The mass dependence enters through the formation redshift $z_f = z_f(M,z)$ that is obtained from the excursion-set relation $\delta_c(z_f) = \delta_c(z) + \tilde w_f \sqrt{\sigma^2(M/2) - \sigma^2(M)}$ with the median weight $\tilde w_f = \sqrt{2\ln(1+\alpha_f)}$, $\alpha_f = 0.815\, e^{-1/4}\, 2^{0.707}$ \cite{Giocoli:2011hz}. The subhalo mass function is shown in the upper panel of Fig.~\ref{fig:shmf}.

The subhalos are distributed within their host following the radial profile motivated by $N$-body results~\cite{Green:2021vtc, Klypin:2010qw}. Subhalos selected by bound mass trace the host density profile in the outskirts but are strongly depleted in the central region by tidal stripping. Following Ref.~\cite{Green:2021vtc} we quantify this through the radial bias function $B_{\rm sub}(x)$ fit to the Bolshoi simulation results~\cite{Klypin:2010qw}: 
\be
    B_{\rm sub}(x) = \left[1+\left(\frac{x}{x_0}\right)^{-5/2}\right]^{-1/2} , \quad x = \frac{r}{r_{200}} \,,
\ee
with $x_0 = 0.86 $. The radius $r_{200}$ is defined such that the mean density within it is 200 times the critical density, with $r_{200} = c_{200} r_s$, where $r_s$ is the NFW scale radius and $c_{200}$ is the concentration parameter. Both refer to the full halo mass $M$ of Eq.~\eqref{eq:reducedhost} rather than to the reduced mass of the smooth component, matching the convention in which the bias function is calibrated. The radial distribution of the subhalos is then $\td N_{\rm sub}/\td r \propto r^2 \rho_{\rm NFW}(r) B_{\rm sub}(r/r_{200})$. The adopted profile is shown in the lower panel of Fig.~\ref{fig:shmf}, where we also compare it with other publicly available substructure models. Each subhalo is projected onto the lens plane of the host halo, and its convergence and shear are evaluated using a circular NFW profile with the same concentration-mass relation evaluated as for field halos.

\begin{figure}
    \centering
    \includegraphics[width=1.0\linewidth]{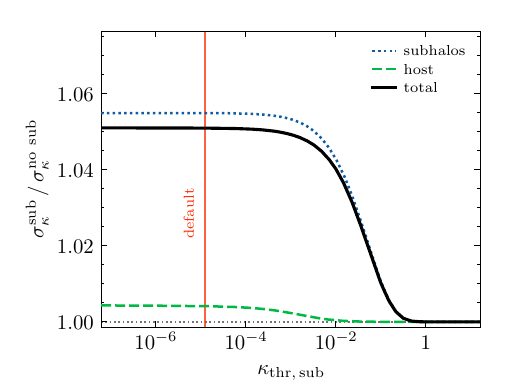}
    \caption{Substructure effect to the convergence standard deviation at $z_s=1$ as a function of the per-subhalo convergence threshold. The vertical line indicates the default threshold used in the simulator.}
    \label{fig:subhalo-kappa}
\end{figure}

Explicitly realizing every subhalo down to the mass floor would be computationally expensive. We therefore render a subhalo only when its convergence exceeds $\kappa_{\rm thr,sub} = 0.1\,\kappa_{\rm thr}$, where $\kappa_{\rm thr}$ is the host-halo threshold defined above. Subhalos below this threshold are not discarded from the mass budget: their mass remains in the smooth host component. Figure~\ref{fig:subhalo-kappa} shows that the adopted threshold lies on the converged plateau of the total substructure contribution to $\sigma_\kappa$. Including substructure increases $\sigma_\kappa$ by approximately $5\%$ at $z_s=1$, while moving from the zero-threshold limit to the adopted threshold changes this contribution by less than $0.1\%$. 

In addition to the threshold on $\kappa$, we impose a minimal mass cut-off of $M_{\rm min} = 10^7 M_\odot$ for both halos and subhalos. We discuss the mass cutoff further in Sec.~\ref{sec:comparison}.

\subsection{Magnification PDF}
\label{sec:pdf}

Following Ref.~\cite{Vaskonen:2026ubd}, we compute the PDF of the magnification $\mu$ by generating Monte Carlo realizations of the structures around the line of sight to a source at redshift $z_s$.\footnote{The stochastic lens model code, extending that of Ref.~\cite{Vaskonen:2026ubd}, is publicly available at \href{https://github.com/vianvask/halos}{https://github.com/vianvask/halos}.} In each realization, we sum the convergences and shears of the large-scale density field, the field halos, the filaments, the sub-threshold background, and the subhalos of each host, and evaluate the magnification using Eq.~\eqref{eq:mueq}. Since the sum of the lensing contributions defines the convergence relative to an empty universe, we subtract the ensemble mean $\langle\kappa\rangle$ from each realization. This procedure yields the magnification PDF in the image plane, $\td P_I/\td\mu$. The corresponding source-plane PDF follows from 
\be
    \frac{\td P_S}{\td\mu} \propto \mu^{-1} \frac{\td P_I}{\td\mu} \,,
\ee
where the factor $\mu^{-1}$ accounts for the change in area between the image and source planes induced by lensing. 

\begin{figure}
    \centering
    \includegraphics[width=\columnwidth]{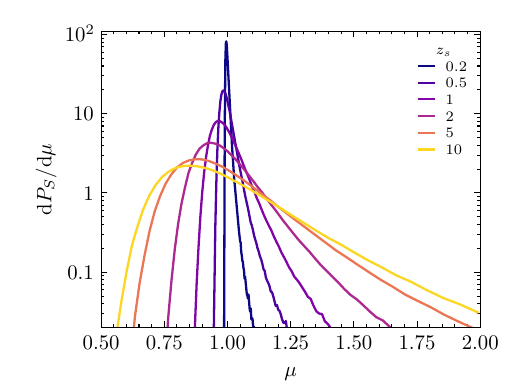}
    \caption{Source-plane magnification PDF computed with the full lens model of this section at the Planck 2018 values from $1.6\times10^6$ realizations per source redshift.}
    \label{fig:magpdf_zs}
\end{figure}

The resulting source plane PDFs for different source redshifts are shown in Fig.~\ref{fig:magpdf_zs} at the Planck 2018 best-fit values~\cite{Planck:2018vyg} (see Table~\ref{tab:param}). The distributions are highly non-Gaussian and get broader with increasing source redshift. As most lines of sight are underdense, while the relatively rare overdense lines of sight can produce large magnifications, the distributions peak below $\mu = 1$ and exhibit a heavy tail towards high magnifications. 

Independently of the detailed properties of the lens population, the magnification PDF has the following structural properties:
\begin{enumerate}[leftmargin=12pt]
    \item Flux conservation~\cite{1976ApJ...208L...1W} fixes the source plane mean magnification $\langle \mu \rangle_S = 1$ since lensing redistributes flux between lines of sight without creating or destroying it.
    
    \item The low-magnification side falls off sharply, and each curve has a lower cutoff corresponding to the minimum matter content along the line of sight. This limiting case is the empty beam~\cite{Dyer1972dis}, in which the beam contains no matter.\footnote{The empty beam bounds $\mu$ from below only while $\kappa<1$. A line of sight passing close to a halo center instead has $\kappa$ well above unity, and once $(1-\kappa)^2-\gamma^2>(1+\langle\kappa\rangle)^2$ the beam is demagnified again, by an excess of matter rather than by its absence. Such core-crossing lines of sight are genuine strong-lensing configurations, for which Eq.~\eqref{eq:mueq} describes only one image of a multiply imaged source. These configurations are, however, very rare and excluded from the emulator training set of Sec.~\ref{sec:ml}.}

    \item The high-magnification side approaches the power law $\td P_S/\td\mu \propto \mu^{-3}$. This behavior arises from fold caustics: near a fold, the cross section for a magnification above $\mu$ scales as $\mu^{-3}$~\cite{Blandford:1986zz}. This asymptotic regime lies beyond the range shown in Fig.~\ref{fig:magpdf_zs}.
\end{enumerate}

\begin{figure}
    \centering
    \includegraphics[width=1.0\columnwidth]{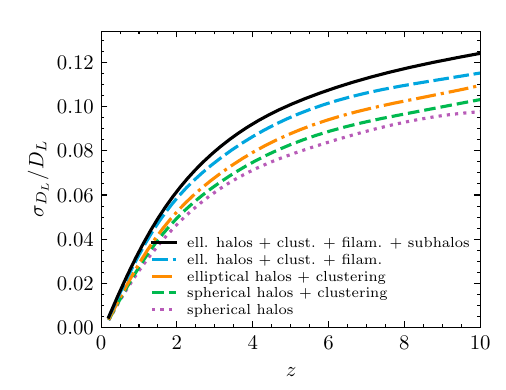}
    \caption{The standard deviation of luminosity distances induced by weak lensing. The curves show, cumulatively, the lens model of Ref.~\cite{Vaskonen:2026ubd}, the effect of adding the subhalo population, and the effect of adding the correlated clustering of field halos.}
    \label{fig:variance_DL}
\end{figure}

Figure~\ref{fig:variance_DL} shows the standard deviation of the luminosity distance, $\sigma_{D_L}$, induced by weak lensing and decomposed into different contributions. Each contribution adds about 10\% to the standard deviation. We compare our result with other works in the literature in Sec.~\ref{sec:comparison}.

\section{Machine learning}
\label{sec:ml}

The Monte Carlo model of Sec.~\ref{sec:lensing} provides an estimate of the magnification PDF, but regenerating it at every point of a parameter scan is computationally expensive. We therefore construct an emulator for the magnification PDF $\td P(\mu|\vect c)/\td \mu$ fitted once to the simulated samples and evaluated thereafter in place of the simulation. The vector $\vect c = (z_s,\vect\Theta)$, which includes the source redshift $z_s$ and the six cosmological parameters of Table~\ref{tab:param}, is called the context. Every quantity the emulator predicts is a function of $\vect c$ alone.

The emulator combines a conditional normalizing flow with explicit physical support and an asymptotic caustic constraint. A conditional Sum-of-Squares Polynomial Flow (SOSPF) with a standard Gaussian base distribution models the bulk of the PDF. A coordinate transformation imposes the adopted empty-beam cutoff $\mu_{\rm cut}$, while an exponential relaxation joins the flow continuously in value and first derivative to a tail whose slope approaches the fold-caustic limit $\td P_S/\td\mu\propto\mu^{-3}$. We describe each component below, writing the flow explicitly as a chain of coordinate transformations, and collect the tail handover in Appendix~\ref{app:emulator}.

\subsection{Training data}

A configuration is one choice of the context $\vect c$, and a ray is one simulated line of sight. Each configuration carries a Monte Carlo sample of magnifications $\mu$ at fixed $\vect c$. The base training set contains 1983 configurations drawn from an optimized Latin Hypercube Sampling (LHS) of the parameter space, totaling approximately $1.19\times10^7$ magnification realizations. To improve coverage near the boundaries of the parameter space, we augment it with 1712 configurations covering all eight corners of the $(h,\Omega_M,\sigma_8)$ parameter volume across eight redshift slices, contributing an additional $1.03\times10^7$ rays. The total training pool thus comprises 3695 configurations and $2.22\times10^7$ rays. The calibrated domain is given in Table~\ref{tab:param}. Predictions outside this domain are extrapolations and are not included in the accuracy statements below. An independent set of 1088 configurations ($6.53\times10^6$ rays) serves as the validation split during training. Model certification and performance benchmarks are evaluated on a completely untouched held-out test pool of 1089 configurations ($6.53\times10^6$ rays), providing the decisive generalization test across the full cosmological volume.

\begin{table}
\centering
\begin{tabular}{lcc}
\hline\hline
Parameter & Range & Planck 2018 \\
\hline
$h$               & $[0.55,\, 0.80]$  & $0.674$ \\
$\Omega_M$        & $[0.15,\, 0.50]$  & $0.315$ \\
$\sigma_8$        & $[0.40,\, 1.50]$  & $0.811$ \\
$\Omega_B$        & $[0.03,\, 0.07]$  & $0.0493$\\
$n_s$             & $[0.94,\, 0.99]$  & $0.965$ \\
$z_{\rm eq}$      & $[3300,\, 3500]$  & $3402$ \\
\hline
$z_s$             & $[0.2,\, 12]$     & $-$     \\
\hline\hline
\end{tabular}
\caption{The calibration ranges of the source redshift $z_s$ and the cosmological parameters and their Planck 2018 best-fit values~\cite{Planck:2018vyg}.}
\label{tab:param}
\end{table}

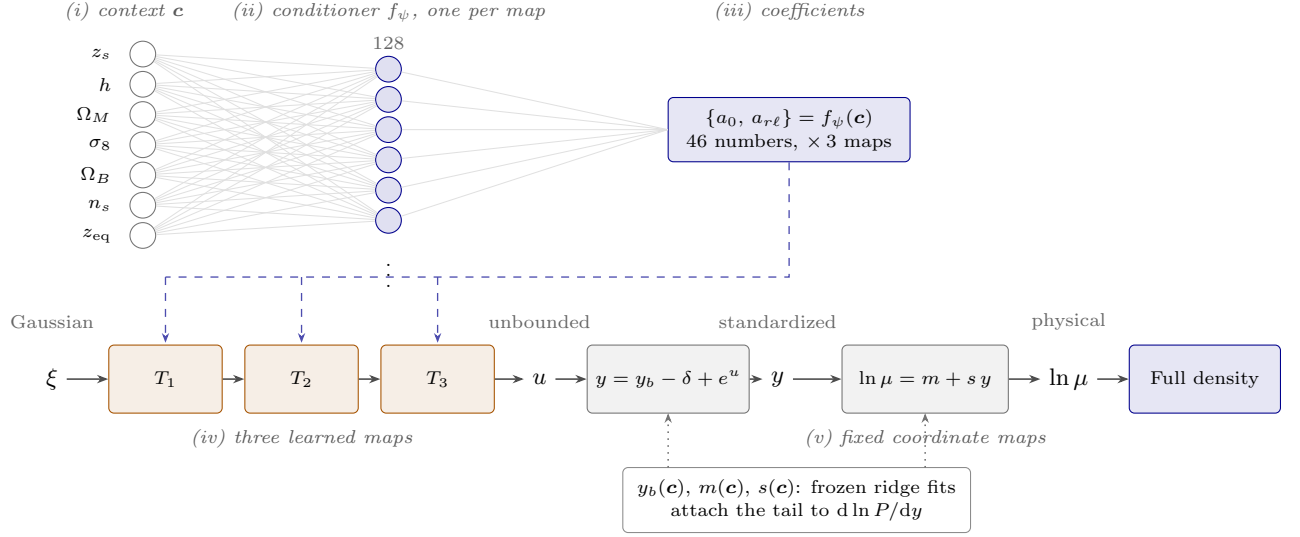
\begin{figure*}
\centering
  \begin{tikzpicture}[
      every node/.style={font=\footnotesize},
      cnode/.style={circle, draw=black!55, fill=white, inner sep=0pt, minimum size=3.4mm},
      hnode/.style={circle, draw=blue!55!black, fill=blue!55!black!12, inner sep=0pt, minimum size=3.4mm},
      clab/.style={font=\scriptsize, anchor=east},
      tbox/.style={rectangle, rounded corners=2pt, draw=orange!65!black, fill=orange!65!black!10,
                   minimum width=1.5cm, minimum height=0.90cm, align=center, font=\scriptsize},
      fbox/.style={rectangle, rounded corners=2pt, draw=black!55, fill=black!5,
                   minimum width=1.9cm, minimum height=0.90cm, align=center, font=\scriptsize},
      cbox/.style={rectangle, rounded corners=2pt, draw=blue!55!black, fill=blue!55!black!10,
                   minimum width=3.2cm, minimum height=0.85cm, align=center, font=\scriptsize},
      tag/.style={font=\scriptsize\itshape, text=black!65},
      elab/.style={font=\scriptsize, text=orange!65!black},
      note/.style={font=\scriptsize, text=black!60},
      wire/.style={draw=black!12, line width=0.25pt},
      rbox/.style={rectangle, rounded corners=2pt, draw=black!45, fill=white,
                 minimum width=3.5cm, minimum height=0.78cm, align=center, font=\scriptsize},
    abox/.style={rectangle, rounded corners=2pt, draw=black!45, fill=white,
                 minimum width=3.4cm, minimum height=0.78cm, align=center, font=\scriptsize},
    ridgearr/.style={draw=black!55, dotted, line width=0.6pt, -{Stealth[length=1.3mm]}},
    feed/.style={draw=blue!55!black!70, dashed, line width=0.5pt},
      feedarr/.style={draw=blue!55!black!70, dashed, line width=0.5pt, -{Stealth[length=1.3mm]}},
      flowarr/.style={draw=black!70, line width=0.6pt, -{Stealth[length=1.5mm]}},
      ruler/.style={draw=black!70, line width=0.5pt},
      tick/.style={draw=black!85, line width=0.7pt},
      warp/.style={draw=orange!65!black!45, line width=0.35pt},
      silh/.style={draw=blue!55!black, line width=0.7pt, fill=blue!55!black!10},
      edge/.style={draw=orange!65!black, line width=0.7pt, dashed}
  ]

  \node[tag] at (1.30,7.66) {(i) context $\vect c$};
  \node[tag] at (4.80,7.66) {(ii) conditioner $f_\psi$, one per map};
  \node[tag] at (10.10,7.66) {(iii) coefficients};

  \foreach \lab [count=\i from 0] in {{$z_s$},{$h$},{$\Omega_M$},{$\sigma_8$},{$\Omega_B$},{$n_s$},{$z_{\rm eq}$}}{
    \node[cnode] (c\i) at (1.55,{7.10-0.40*\i}) {};
    \node[clab] at (1.25,{7.10-0.40*\i}) {\lab};}
  \foreach \i in {0,...,5} \node[hnode] (ha\i) at (4.80,{6.90-0.40*\i}) {};
  \node[font=\scriptsize] at (4.80,4.28) {$\vdots$};
  \node[note] at (4.80,7.24) {$128$};

  \foreach \i in {0,...,6} \foreach \j in {0,...,5} \draw[wire] (c\i) -- (ha\j);

  \node[cbox] (coef) at (10.10,6.10) {$\{a_0,\,a_{r\ell}\}=f_\psi(\vect c)$\\[0.3mm]
    $46$ numbers, $\times\,3$ maps};
  \foreach \i in {0,...,5} \draw[wire] (ha\i) -- (coef.west);

  \node[tag] at (3.65,2.02) {(iv) three learned maps};
  \node[tag] at (11.90,2.02) {(v) fixed coordinate maps};

  \node[note] at (0.35,3.56) {Gaussian};
  \node[note] at (6.80,3.56) {unbounded};
  \node[note] at (9.95,3.56) {standardized};
  \node[note] at (13.80,3.56) {physical};
  \node[font=\small] (xin) at (0.35,2.80) {$\xi$};
  \node[tbox] (t1) at (1.85,2.80) {$T_1$};
  \node[tbox] (t2) at (3.65,2.80) {$T_2$};
  \node[tbox] (t3) at (5.45,2.80) {$T_3$};
  \node[font=\small] (uu) at (6.80,2.80) {$u$};
  \node[fbox] (bd) at (8.50,2.80) {$y=y_b-\delta+e^{u}$};
  \node[font=\small] (yy) at (9.95,2.80) {$y$};
  \node[fbox, minimum width=2.2cm] (ls) at (11.90,2.80) {$\ln\mu=m+s\,y$};
  \node[font=\small] (mm) at (13.80,2.80) {$\ln\mu$};
  \node[fbox, draw=blue!55!black, fill=blue!55!black!10, minimum width=2.0cm] (out) at (15.60,2.80)
    {Full density};

  \draw[flowarr] (xin) -- (t1);  \draw[flowarr] (t1) -- (t2);
  \draw[flowarr] (t2) -- (t3);   \draw[flowarr] (t3) -- (uu);
  \draw[flowarr] (uu) -- (bd);   \draw[flowarr] (bd) -- (yy);
  \draw[flowarr] (yy) -- (ls);   \draw[flowarr] (ls) -- (mm);
  \draw[flowarr] (mm) -- (out);

  \draw[feed] (coef.south) -- (10.10,4.15) -- (1.85,4.15);
  \foreach \x in {1.85,3.65,5.45} \draw[feedarr] (\x,4.15) -- (\x,3.28);
  \node[rbox, minimum width=4.6cm, minimum height=0.86cm] (aux) at (10.20,1.20)
    {$y_b(\vect c)$, $m(\vect c)$, $s(\vect c)$: frozen ridge fits\\[0.3mm]
     attach the tail to $\td\ln P/\td y$};
  \draw[ridgearr] (8.50,1.63) -- (8.50,2.35);
  \draw[ridgearr] (11.90,1.63) -- (11.90,2.35);

    \end{tikzpicture}
    \caption{Architecture of the conditional magnification emulator. The context $\vect c$ (i) is the only input. The conditioner (ii) turns it into the coefficients (iii) of the three monotonic maps (iv) of Eq.~\eqref{eq:sospf}, which deform a standard Gaussian variable $\xi$ into the coordinate $u$. No map coefficients are stored independently: at each context they are generated by the corresponding conditioner, one per map and drawn only once here. The two maps (v) are fixed before fitting and carry $u$ to the standardized coordinate $y$ of Eq.~\eqref{eq:flow-standardization} and then to $\ln\mu$. Their $y_b$, $m$ and $s$ also depend on the context through frozen ridge fits made at an earlier stage, which is why they are drawn as a separate input. The asymptotic relaxation bridge of Appendix~\ref{app:emulator} is attached to $\td\ln P/\td y$ at this standardized stage, before the map to $\ln\mu$; only the unit-flux shift acts on the physical variable directly. Figure~\ref{fig:emulator-components} shows the deformation produced by the fitted maps, stage by stage, at the Planck 2018 cosmology.}
    \label{fig:emulator_arch}
\end{figure*}

\begin{figure*}
    \centering
    \includegraphics[width=\linewidth]{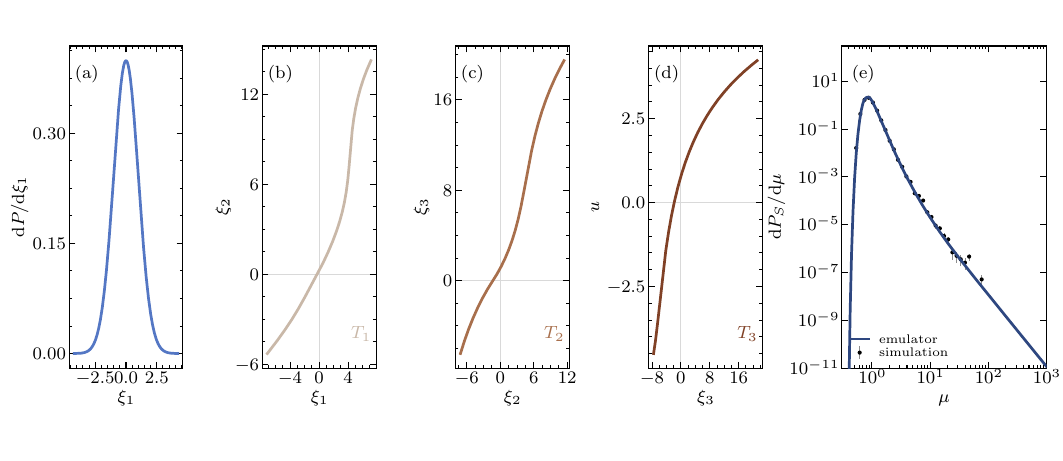}
    \caption{The flow shown as the sequence of maps that produces it, at the Planck 2018 cosmology and $z_s=10$. Panel (a) is the standard Gaussian $\xi$ the chain starts from. Panels (b)--(d) plot the three monotonic maps $T_1$, $T_2$ and $T_3$ of Eq.~\eqref{eq:sospf}. Their composition carries $\xi$ to the coordinate $u$ of Eq.~\eqref{eq:boundary-transform}, in which the flow is fitted. Panel (e) is the fully assembled density body against simulation points.}
    \label{fig:emulator-components}
\end{figure*}

\subsection{Conditional normalizing flow}

Figure~\ref{fig:emulator_arch} shows the emulator as a transformation chain leading from a latent Gaussian variable to physical magnification. We describe its stages in turn, in the order the data traverse them, and refer to the labels of that figure throughout. Everything that is fitted to data sits in stages (ii) to (iv), and the two coordinate changes (v) that follow are fixed before that fit begins. The chain carries four names for one random quantity: the Gaussian $\xi$, the unbounded coordinate $u$ in which the flow is fitted, the standardized coordinate $y$, and the physical variable $\ln\mu$. Only the last is observable, and the others are introduced because the distribution takes a simpler form in those coordinates. Each arrow in the transformation chain is an invertible change of variables, so a density known in any one coordinate determines it in all others.

The width of the PDF varies by nearly three orders of magnitude across the calibrated domain, from $\sigma(\ln\mu)\simeq1.2\times10^{-3}$ at $z_s=0.2$ in the least structured cosmologies to $\sigma(\ln\mu)\simeq1$ at $z_s=12$ in the most structured ones, so no single grid in $\ln\mu$ resolves both regimes. Following the location-scale parameterization standard in conditional density estimation and in affine normalizing flows~\cite{Papamakarios:2019fms, Papamakarios:2017tec}, we absorb this variation into the coordinate
\be \label{eq:flow-standardization}
    y = \frac{\ln\mu - m(\vect c)}{s(\vect c)} \,,
\ee
where $m(\vect c)$ is the median of $\ln \mu$ and $s(\vect c)$ is half of its $15.87-84.13\%$ range (standard deviation for Gaussian). Inverted, this is the last stage (v) of Fig.~\ref{fig:emulator_arch}. Both are obtained before training by ridge regression,\footnote{These regressions act as simple conditioning functions, analogous to the neural conditioners in stage~(ii) of Fig.~\ref{fig:emulator_arch}, but determine the coordinate transformations.} and are held fixed while the flow is fitted. The densities are related by the change of variables
\be \label{eq:flow-jacobian}
     \frac{\td P_I(\mu\mid\vect c)}{\td\mu} = \frac{1}{s(\vect c)\,\mu}\,\frac{\td P(y\mid\vect c)}{\td y} \,.
\ee

The flow resolves the physical lower support explicitly. The empty-beam boundary is fixed by the mean unshifted convergence $\bar\kappa(\vect c)$, which is for a line of sight containing no matter, $\gamma=0$ and $\kappa = -\bar\kappa(\vect c)$, so Eq.~\eqref{eq:mueq} gives the exact cutoff $\mu_{\rm cut} = 1/[1+\bar\kappa(\vect c)]^2$. A smooth polynomial regression predicts the location $y_b(\vect c)$ of the empty-beam boundary in standardized coordinate $y$. To avoid numerical issues near the boundary, we introduce a fixed offset $\delta=0.05$ \footnote{We have verified that varying $\delta$ over the range $[0.001,0.8]$ changes the predicted PDFs by less than the variation between different training runs.} and fit the flow in the coordinate
\be \label{eq:boundary-transform}
    u = \ln\!\left[y - y_b(\vect c) + \delta\right] , \qquad y > y_b(\vect c) - \delta \,,
\ee
which maps the semi-infinite support smoothly onto the entire real line. Inverting Eq.~\eqref{eq:boundary-transform} yields the Jacobian
\be
 \frac{\td P(y\mid\vect c)}{\td y} = \frac{1}{y - y_b(\vect c) + \delta}\,\frac{\td P(u\mid\vect c)}{\td u} \,,
\ee
so that $u$ is now fixed, by construction, as the coordinate in which we fit the flow, and the two maps in stage (v) of Fig.~\ref{fig:emulator_arch} carry it to $\ln\mu$. Below the modeled empty-beam cutoff $\mu < \mu_{\rm cut} = \exp\!\left[m + s(y_b - \delta)\right]$, the PDF vanishes.

A normalizing flow is a change of variables built from a chain of fitted, monotonic maps. We construct the SOSPF from the transformations~\cite{jaini2019sum}
\be \label{eq:sospf}
    T_J(\xi_J) = a_{J,0} + \int_0^{\xi_J}\!\td t \left[\epsilon + \sum_{r=1}^{R}\left(\sum_{\ell=0}^{L}a_{J,r\ell}\,t^{\ell}\right)^{\!2}\right] \,,
\ee
with $R=5$ polynomials of degree $L=8$ and a minimum slope $\epsilon=10^{-3}$. 
We compose three such maps, $J=1,2,3$, writing $\xi_1\equiv\xi$ for a standard Gaussian variable, $\xi_{J+1}\equiv T_J(\xi_J)$ and $\xi_4\equiv u$ for the coordinate in which the flow is ultimately fitted. The transformation gives the distribution of $\xi_{J+1}$ as
\be \label{eq:flow-cov}
    \frac{\td P_{J+1}}{\td\xi_{J+1}} = \frac{\td T_J^{-1}(\xi_{J+1})}{\td\xi_{J+1}} \frac{\td P_J}{\td\xi_J}\bigg|_{\xi_J=T_J^{-1}(\xi_{J+1})} \,.
\ee
Composing Eq.~\eqref{eq:flow-cov} over $J=1,2,3$ multiplies the three Jacobians and gives the density of $u$ directly in terms of the initial Gaussian distribution. 

A normalizing flow is precisely this construction with each $T_J$ fitted to data rather than prescribed. Since every $T_J$ is monotonic, its inverse exists everywhere, so each factor in Eq.~\eqref{eq:flow-cov} is positive and every stage integrates to unity, for any fitted $T_J$. The Gaussian from which the chain starts is symmetric, unbounded, and falls faster than any power, whereas the magnification PDF is skewed, bounded from below and has a power-law tail. All three features are produced by the composed deformation rather than by the base distribution. Figure~\ref{fig:emulator-components} shows this deformation directly, one map at a time.

The coefficients of each map carry the entire dependence on cosmology,
\be \label{eq:conditioner}
    \{a_{J,0},\,a_{J,r\ell}\} = f_{\psi_J}(\vect c) \,,
\ee
where $f_{\psi_J}$ is the conditioner (ii) of Fig.~\ref{fig:emulator_arch} for map $J$, a feed-forward network with one hidden layer of $128$ units evaluated on the context standardized to zero mean and unit variance over the training set. Each map needs $46$ coefficients, one for each of the $R(L+1)$ polynomial coefficients and one for the integration constant. The three maps are composed and their Jacobians multiply. The weights $\psi_J$ are shared by all contexts, so the emulator interpolates in cosmology through the conditioners rather than by storing a separate fit per configuration. The flow contains only $20\,874$ trainable parameters, obtained by maximizing the likelihood of the training rays with the Adam optimizer \cite{Kingma:2014vow}. 

The parameters $\psi$ of the conditioner network $f_\psi(\vect c)$ in Eq.~\eqref{eq:conditioner} are optimized by minimizing the negative log-likelihood of the training rays~\cite{Papamakarios:2019fms, Dai:2023lcb}:
\be
    \mathcal{L}(\psi) = -\frac{1}{N}\sum_{i=1}^{N}\ln\frac{\td P(u_i\mid\vect c_i)}{\td u} \,.
\ee
The standardization parameters $m(\vect c)$ and $s(\vect c)$ of Eq.~\eqref{eq:flow-standardization} as well as the empty-beam boundary $y_b(\vect c)$ of Eq.~\eqref{eq:boundary-transform} are obtained prior to flow optimization via polynomial ridge regressions and held strictly frozen during training. Rays falling at or below the physical boundary $y_b$ are removed before fitting, which excludes less than $3\times10^{-5}$ of the training sample.

\subsection{High-magnification tail and physical calibration}
\label{sec:tail}

Maximum-likelihood training alone does not determine the high-magnification tail reliably, because only a tiny fraction of simulated rays reaches that regime. The leading fold-caustic gives the asymptote $\td P_I/\td\mu\propto\mu^{-2}$. In the standardized coordinate $y$ this fixes the asymptotic log-density slope to $\lim_{y\to\infty}\td\ln P/\td y=-s(\vect c)$, whose subleading correction is derived in Appendix~\ref{app:emulator}.

We anchor the tail transition at $y = y_0 = 10$, at $10$ distribution widths above the median of $\ln \mu$. To join the flow smoothly to the caustic asymptote without creating derivative kinks or artificial plateaus, we relax the log-density slope exponentially from its flow value $d_0(\vect c) = \td\ln P_{\rm flow}(y\mid\vect c)/\td y|_{y_0}$ to the universal fold-caustic slope $-s(\vect c)$ for all $y \ge y_0$:
\bea \label{eq:tail-relaxation}
    \ln P(y\mid\vect c) &= f_0 - s(\vect c)\,(y - y_0) \\
    &\qquad + h\left[d_0 + s(\vect c)\right]\left[1 - e^{-(y - y_0)/h}\right] \,,
\eea
with characteristic relaxation scale $h(\vect c) = 1/s(\vect c)$ and $f_0(\vect c) = \ln P_{\rm flow}(y_0\mid\vect c)$. In physical magnification $\mu$, this exponential relaxation in $y$ corresponds to an approach at rate $\mathcal{O}(1/\mu)$, exactly mirroring the universal subleading correction derived in Appendix~\ref{app:emulator}. The resulting composite density is strictly positive, continuously differentiable across $y_0$.

After joining the flow to the asymptotic tail, we normalize the composite density and calibrate it to satisfy flux conservation, $\langle\mu\rangle_S=1$. For the normalized image-plane density, this is equivalent to $\langle\mu^{-1}\rangle_I=1$ under the source-plane reweighting described in Sec.~\ref{sec:pdf}. The composite comes close to this on its own, satisfying it to better than one percent, though the residual grows with source redshift. We close the gap by adapting the magnification rescaling of Ref.~\cite{Turker:2025rdt}, which brings $\langle\mu^{-1}\rangle_I$ to unity to better than $10^{-4}$ for every held-out configuration.

\begin{figure}
    \centering
    \includegraphics[width=\linewidth]{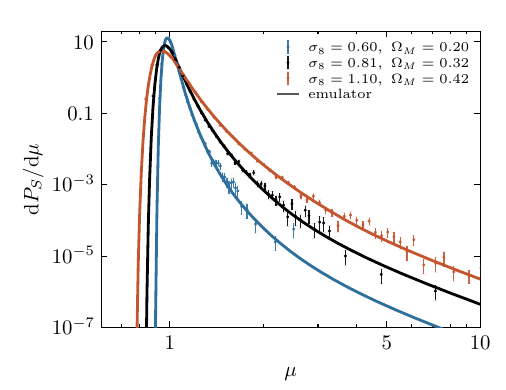}
    \caption{Source-plane magnification distributions $\td P_S/\td \mu$ predicted by the emulator (solid lines) compared with simulation (points) at $z_s = 1$ across three benchmark cosmologies, with the remaining cosmological parameters fixed to their Planck~2018 values.}
    \label{fig:emulator-benchmark}
\end{figure}

\begin{figure}
    \centering
    \includegraphics[width=\linewidth]{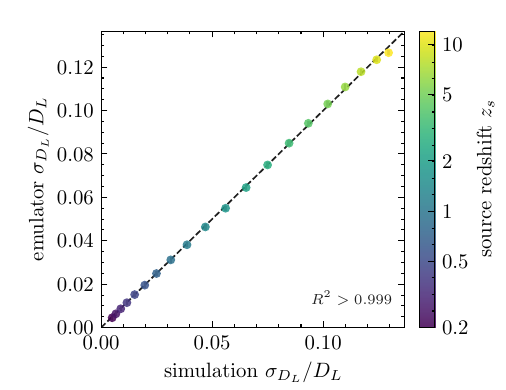}
    \caption{Emulated versus simulated standard deviation of the luminosity distance at the Planck 2018 cosmology, using 20 redshifts with $2\times10^4$ realizations each.}
    \label{fig:sigma_mu}
\end{figure}

Figure~\ref{fig:emulator-benchmark} compares the resulting source-plane PDFs with Monte Carlo simulations at $z_s=1$ for three benchmark cosmologies. These span increasingly structured universes, with both the matter abundance $\Omega_M$ and the clustering amplitude $\sigma_8$ increasing. As these parameters increase, lensing produces a broader range of magnifications. Consequently, the distribution broadens, its peak shifts toward lower magnifications, and the high-magnification tail becomes increasingly prominent. The emulator captures this evolution, reproducing the peak and asymmetric body of the PDF while providing a smooth continuation into the sparsely sampled tail. The agreement is closest in the well-populated body. Figure~\ref{fig:sigma_mu} provides a complementary check through the relative luminosity distance scatter, $\sigma_{D_L}/D_L$, at the Planck 2018 cosmology. The emulator follows its increase over $0.2\leq z_s\leq12$, with a median absolute fractional difference of approximately $0.9\%$ across the 20 redshifts.

\subsection{Validation and held-out accuracy}
\label{sec:validation}

We evaluate the emulator across the complete, untouched held-out test split of 1089 configurations ($6.53\times10^6$ rays). To provide a comprehensive assessment that does not rely on any single binning choice, we measure accuracy with two statistical metrics.

First, following the approach used in Ref.~\cite{Turker:2025rdt}, we evaluate the Kullback--Leibler divergence on a width-relative standardized grid. The range $-6<y<12$ is divided into 120 equal bins of width $0.15 s(\vect c)$, bins containing fewer than five rays are excluded, and the discrete probabilities are formed via continuous cell quadrature:
\be \label{eq:kl-y}
    KL = \sum_i q_i\ln\!\left(\frac{q_i}{p_i}\right) .
\ee
Across all 1089 test configurations, we find the median $KL = 0.0048$.

Second, to eliminate any potential sensitivity to tail binning or Poisson noise, we compute the equal-mass adaptive quantile divergence $KL_{\rm quant}$ ($K=50$ bins, each containing exactly the same number of rays), incorporating the finite-sample Miller--Madow correction. We get the median debiased $KL_{\rm quant} = 0.00153$, demonstrating that the small residual in fixed binning is dominated by sparse tail noise~\cite{Paninski:2003est}.

\section{Comparison with earlier works}
\label{sec:comparison}

We compare the present framework with previous work by examining the weak-lensing-induced scatter in the luminosity distance, the shape and width of the predicted magnification PDFs, and the internal accuracy of the emulators. These comparisons are intended as diagnostics of modeling choices rather than as a benchmark, since the different studies use distinct forward models and validation samples.

\begin{figure}
    \centering
    \includegraphics[width=\linewidth]{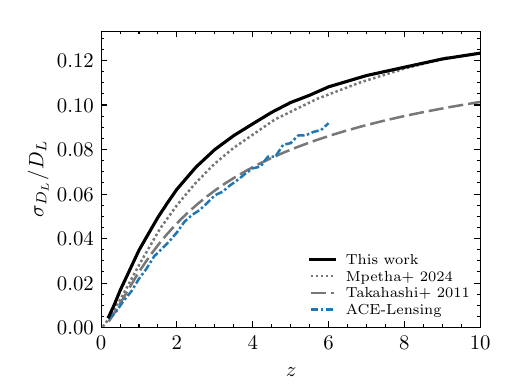}
    \caption{Comparison of the standard deviation of luminosity distances induced by weak lensing.}
    \label{fig:comparison-variance}
\end{figure}

\begin{figure}
    \centering
    \includegraphics[width=0.96\linewidth]{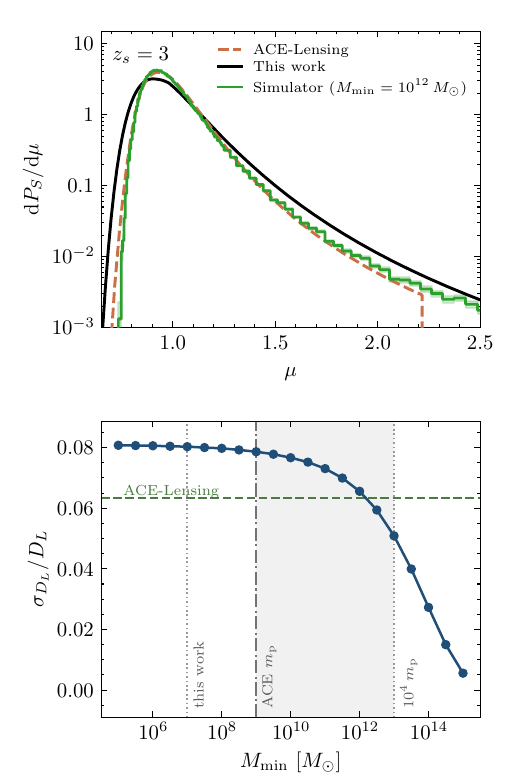}
    \caption{\textbf{Top:} Source-plane magnification PDF at $z_s=3$ and the Planck 2018 cosmology. The restricted-resolution simulator tracks \texttt{ACE-Lensing} closely, while the fully resolved simulator predicts a broader peak and a heavier tail. \textbf{Bottom:} Standard deviation of the luminosity distance at $z_s=3$ as the mass floor of our stochastic model is varied.}
    \label{fig:ace-mmin-comparison}
\end{figure}

We start with the standard deviation of the luminosity distance, $\sigma_{D_L}/D_L$, shown in Fig.~\ref{fig:comparison-variance}. The full stochastic model agrees well with the $\Lambda$CDM curve of Ref.~\cite{Mpetha:2024xiu}, which is not a simulation result but the halo-model prediction $\sigma_{D_L}/D_L \simeq \sigma_\kappa$, obtained by integrating the convergence power spectrum up to $\ell_{\max}=10^7$ with the nonlinear matter power spectrum of \texttt{HMcode-2020}~\cite{Mead:2020vgs} following Ref.~\cite{Martinelli:2022elq}. This agreement is expected, as the one-halo and two-halo terms of \texttt{HMcode-2020} carry the same physical content as our lens population, and it serves as a consistency check of the body of our PDF. The $N$-body-based results of Takahashi et al.~\cite{Takahashi:2011qd} and \texttt{ACE-Lensing}~\cite{Turker:2025rdt} both lie significantly below our result, for different reasons. \texttt{ACE-Lensing} is based on simulations in a large box, $\mathcal{O}(100\,{\rm Mpc})$, but with a particle mass of $\mathcal{O}(10^9 M_\odot)$ and a $\mathcal{O}(10\,{\rm kpc})$ projection grid. Halos below $\mathcal{O}(10^{12}\Msun)$ are consequently unresolved and the inner profiles of more massive halos are smoothed. We show in Fig.~\ref{fig:ace-mmin-comparison} that imposing a corresponding mass floor in our model reproduces the magnification PDF of \texttt{ACE-Lensing}. Takahashi et al. instead reach a particle mass of $\mathcal{O}(10^7 M_\odot)$ and a $\mathcal{O}(1\,{\rm kpc})$ grid, but in a small $\mathcal{O}(10\, {\rm Mpc})$ box that contains of order one halo above $10^{14}\,M_\odot$ and that lacks all density modes longer than the box size. This removes both the rare massive lenses and the long-wavelength clustering of the lens population. Ref.~\cite{Mpetha:2024xiu} reached the same conclusion, showing that the Takahashi et al.\ result is recovered from the halo-model integral once the $\ell_{\min}$ and $\ell_{\max}$ implied by their box and grid are imposed.

\texttt{ACE-Lensing}~\cite{Turker:2025rdt} emulates source-plane magnification PDFs constructed from a suite of 70 dark-matter-only $N$-body simulations. Its parameter space is four dimensional, $(\Omega_M,\sigma_8,w,h)$, its source redshift range is $z_s\in[0.2,6]$, and its PDFs are tabulated over $0.1<\mu<6$. Our emulator instead conditions on six cosmological parameters over the wider ranges, extends to $z_s=12$, and covers the full support of $\mu$ down to the empty-beam cutoff and through the caustic tail. In terms of internal accuracy, \texttt{ACE-Lensing} reports a median KL divergence of $KL = 0.0069$ between emulated and simulated PDFs on 290 test configurations, whereas our conditional flow reaches a median $KL = 0.0048$ on 1089 held-out configurations when evaluated on the same width-relative binning (Sec.~\ref{sec:validation}). The two values indicate comparable interpolation accuracy, but their ratio should not be read as a fractional improvement, since each is measured against a different simulator and test set, with different effective support.

While \texttt{ACE-Lensing} inherits the self-consistent nonlinear density field of its $N$-body simulations, it is necessarily limited by their particle mass and grid resolution. Our stochastic construction instead makes the lens population explicit, so that the halo mass floor $M_{\rm min}$ can be varied directly. The top panel of Fig.~\ref{fig:ace-mmin-comparison} compares the source-plane PDF at $z_s=3$ with the \texttt{ACE-Lensing} prediction. Raising the mass floor of our model to $M_{\rm min}=10^{12}\,\Msun$ reproduces the \texttt{ACE-Lensing} PDF closely, whereas the production choice $M_{\rm min}=10^7\,\Msun$ yields a broader peak and a heavier tail. This is consistent with the resolution of their simulations: with a particle mass of $\mathcal{O}(10^{9})\,\Msun$, halos are reliably resolved only above $\mathcal{O}(10^{3})$ particles, i.e.\ $M\gtrsim10^{12}\,\Msun$. The bottom panel makes this quantitative through the scatter in the luminosity distance as a function of $M_{\rm min}$. At $z_s=3$, $\sigma_{D_L}/D_L$ rises from $\approx0.006$ at $M_{\rm min}=10^{15}\,\Msun$ to $\approx0.08$ at $M_{\rm min}=10^{7}\,\Msun$, with the increase spread over many decades in mass, and changes by less than one percent when the floor is lowered further, so that our production result is converged with respect to the mass floor. The value obtained from the \texttt{ACE-Lensing} PDF, $\sigma_{D_L}/D_L\approx0.063$, crosses our curve at $M_{\rm min}\approx10^{12}\,\Msun$, in agreement with the resolution estimate above.

\section{Conclusions}
\label{sec:conclusions}

We have improved the stochastic weak-lensing model in two ways. First, every lensing halo is now split into a smooth component and a population of subhalos drawn from the evolved subhalo mass function, distributed according to the radial profile measured in $N$-body simulations, with the total halo mass conserved. Second, the independent Poisson realization of the lens population is replaced by counts modulated by a stochastic realization of the long-wavelength density field along the line of sight, with the sub-threshold background drawn conditionally on the same realization so that resolved lenses and the unresolved background fluctuate coherently. Both extensions broaden the magnification PDF and its variance agrees with the halo-model obtained from the nonlinear matter power spectrum in~\cite{Mpetha:2024xiu}. The $N$-body-based estimates in the literature lie below it. We trace this to the limited mass resolution of the $N$-body simulations underlying \texttt{ACE-Lensing}~\cite{Turker:2025rdt} and to the small simulation volume of Takahashi et al.~\cite{Takahashi:2011qd}. In particular, imposing a mass floor of $10^{12}\,M_\odot$ in our model reproduces the \texttt{ACE-Lensing} PDF, whereas our production result is converged with respect to the adopted floor of $10^7\,M_\odot$.

Building on this model, we have constructed \texttt{FLUMEN}, a conditional normalizing-flow emulator of the magnification PDF as a function of the source redshift $z_s\le12$ and the six cosmological parameters $\{h,\Omega_M,\Omega_B,\sigma_8,n_s,z_{\rm eq}\}$. A sum-of-squares polynomial flow with a context-dependent conditioner models the body of the PDF in a standardized coordinate, while the physical properties of the PDF that a purely data-driven fit would miss are built in analytically: the empty-beam cutoff at the low-magnification end and the fold-caustic power-law tail at the high-magnification end, joined to the flow by a smooth relaxation. The emulator reproduces the simulated PDFs on held-out configurations with a median KL divergence of $0.0048$ at a negligible cost per evaluation compared to the underlying Monte Carlo. This makes likelihood-based inference from the full magnification distribution, of the kind proposed in Ref.~\cite{Vaskonen:2026ubd} for GW standard sirens, practical over the full cosmological parameter space, and equally applicable to supernova and quasar samples.

\section*{Acknowledgments} 
We thank Juan Urrutia for useful discussions on weak gravitational lensing, and Joosep Pata and Martin Vasar for useful discussions related to the development of the emulator. This work was supported by the Estonian Research Council grants TARISTU24-TK3 and TARISTU24-TK10, and the Center of Excellence program TK202 of the Estonian Ministry of Education and Research.

\appendix

\section{Asymptotic caustic tail}
\label{app:emulator}

In the extreme high-magnification regime, ray-tracing statistics become sparse, but general lensing theory fixes the functional form. Expanding at large $\mu$ we get
\be \label{eq:caustic-expansion}
    \frac{\td P_I}{\td\mu}(\mu\mid\vect c) = C_1(\vect c)\,\mu^{-2} + C_2(\vect c)\,\mu^{-3} + \mathcal{O}(\mu^{-4}) \,,
\ee
whose leading exponent is the universal fold-caustic value~\cite{Blandford:1986zz}. Converting Eq.~\eqref{eq:caustic-expansion} to the logarithmic density gives
\be
    \ln\frac{\td P_I}{\td\ln\mu} = \ln C_1(\vect c) - \ln\mu + \frac{C_2(\vect c)}{C_1(\vect c)} e^{-\ln\mu} + \mathcal{O}(e^{-2\ln\mu}) \,,
\ee
so that its logarithmic slope is
\be \label{eq:logslope}
    \frac{\td\ln\td P_I/\td\ln\mu}{\td\ln\mu} = -1 - \frac{C_2(\vect c)}{C_1(\vect c)} \mu^{-1} + \mathcal{O}(\mu^{-2}) \,.
\ee
The asymptotic slope is exactly $-1$, and if $C_2$ is indeed the leading correction, the departure from it decays as $1/\mu$.

Rather than imposing an artificial piecewise cubic bridge on a finite interval, we smoothly relax the log-density slope from its flow value $d_0 = \td\ln P_{\rm flow}/\td y|_{y_0}$ to the caustic asymptote $-s(\vect c)$ via
\be
 \frac{\td\ln P(y\mid\vect c)}{\td y} = -s(\vect c) + \left[d_0 + s(\vect c)\right] e^{-(y - y_0)/h} \,.
\ee
In physical magnification $\mu = \exp[m(\vect c) + s(\vect c)\,y]$, the exponential factor satisfies:
\be
 e^{-(y - y_0)/h} = \exp\!\left[-\frac{\ln(\mu/\mu_0)}{s(\vect c)\,h}\right] = \left(\frac{\mu_0}{\mu}\right)^{\frac{1}{s(\vect c)\,h}} ,
\ee
where $\mu_0 = \exp[m(\vect c) + s(\vect c)\,y_0]$. Together with the leading correction in Eq.~\eqref{eq:logslope}, this justifies our choice of the relaxation scale $h = 1/s(\vect c)$ in Sec.~\ref{sec:tail}.

We note, however, that this subleading term is not universal. Depending on the lens geometry, the analogous correction can instead cancel~\cite{Keeton:2005ad,Alexandrov:2010ns}, pushing the leading correction to $\mathcal{O}(\mu^{-4})$ and predicting a faster relaxation. Cusps, a different type of caustic, contribute at $\mu^{-5/2}$ in the same convention~\cite{Schneider:1992grv,Aazami:2019sjg}, a rate that sits between the two terms of Eq.~\eqref{eq:caustic-expansion}. They would instead slow the relaxation, giving an $e$-folding constant of $2$. 

\bibliography{refs}

\end{document}